\documentclass[pre,aps,reprint,noshowpacs,superscriptaddress,floatfix,letterpaper]{revtex4-2}
\usepackage{amsmath,amssymb,amsbsy,amsfonts,amsthm,bbm,bm,mathtools,mathrsfs}
\usepackage{color}
\usepackage{physics}
\usepackage[dvipsnames]{xcolor}
\pagecolor{white}
\definecolor{LapisLazuli}{RGB}{47, 102, 169}
\usepackage[hypertexnames=false,colorlinks=true,citecolor=LapisLazuli,linkcolor=LapisLazuli,urlcolor=LapisLazuli]{hyperref}
\usepackage{pgfplots}
\pgfplotsset{compat = newest}
\usepackage{tikz}
\usepackage{tcolorbox}
\usepackage{tikz}
\usetikzlibrary{decorations}
\usetikzlibrary{decorations.pathreplacing}

\usepackage{epsfig}
\usepackage{subfigure} 
\usepackage{natbib} 
\usepackage{graphicx} 
\usepackage{mathrsfs}
\usepackage{commath} 

\newcommand{\speedtime}{\tau_{\!\scriptscriptstyle\mathcal{A}}}
\newcommand{\speedtimeheat}{\tau_{\!\scriptscriptstyle\mathcal{Q}}}
\newcommand{\speedtimework}{\tau_{\!\scriptscriptstyle\mathcal{W}}}
\newcommand{\speedtimeenergy}{\tau_{\!\scriptscriptstyle U}}
\newcommand{\speedtimemcx}{\tau_{\!\scriptscriptstyle \mathcal{X}}}
\newcommand{\speedtimex}{\tau_{\!\scriptscriptstyle X}}
\newcommand{\Qdot}{\dot{\mathcal{Q}}}
\newcommand{\Wdot}{\dot{\mathcal{W}}}
\newcommand{\Q}{\mathcal{Q}}
\newcommand{\W}{\mathcal{W}}

\newcommand{\speedtimeTSe}{\tau_{\!\scriptscriptstyle TS_e}}  

\newcommand{\cov}{\operatorname{cov}}
\newcommand{\Hquad}{\hspace{0.5em}}

\newcommand{\jrg}[1]{\textcolor{black}{#1}}

\newcommand{\Intd}{{\rm{d}}}

\usepackage{gensymb} 
\usepackage{mathtools} 
\usepackage{enumitem}
\usepackage{tcolorbox} 

\begin{document}

\title{Thermodynamic speed limits for mechanical work}

\author{Erez~Aghion}
\author{Jason~R.~Green}
\email{jason.green@umb.edu}
\affiliation{Department of Chemistry,\
  University of Massachusetts Boston,\
  Boston, MA 02125
}
\affiliation{Department of Physics,\
  University of Massachusetts Boston,\
  Boston, MA 02125
}


\begin{abstract}

Thermodynamic speed limits are a set of classical uncertainty relations that, so far, place global bounds on the stochastic dissipation of energy as heat and the production of entropy.
Here, instead of constraints on these thermodynamic costs, we derive integral speed limits that are upper and lower bounds on a thermodynamic benefit---the minimum time for an amount of mechanical work to be done on or by a system.
In the short time limit, we \jrg{show how this extrinsic timescale relates to} an intrinsic timescale for work, recovering the intrinsic timescales in differential speed limits from these integral speed limits and turning the first law of stochastic thermodynamics into a first law of speeds.
As physical examples, we consider the work done by a flashing Brownian ratchet and the work done on a particle in a potential well subject to external driving.

\end{abstract}

\date{\today}
\maketitle

Bounds on the efficient generation of mechanical work have been a defining feature of thermodynamics since its inception~\cite{carnot1824reflections,callenThermodynamicsIntroductionThermostatistics1985}.
Real processes fall short of these limits, particularly when the system is small, subject to strong fluctuations, or evolves on timescales that are fast compared to intrinsic timescales~\cite{jarzynskiEqualitiesInequalitiesIrreversibility2011}.
\jrg{Biological systems must often overcome these challenges to generate work on particular timescales; for example, timely motion powered by motor proteins supports} cellular functions and even cell survival~\cite{miller2012molecular,milo2015cell}. 
What limits the energy efficiency of machines\jrg{, both biological and synthetic,} at the nanoscale is that the work done by molecular motions is comparable to the entropic losses through fluctuating, dissipative processes~\cite{bo2014entropy,amano2022insights}.
In efforts to create synthetic molecular motors and optimize their efficiency, a possible workaround would be to design their structure to generate power more quickly than entropic losses.
That is, we might be able to engineer energy-efficient systems with a better understanding of the trade-off between the relative speed of work generation and dissipation in stochastic processes~\cite{Bone2022a,Bone2022b}.

Thermodynamic speed limits~\cite{JRGReview2022} have the potential to make new inroads into this challenge. 
Like quantum speed limits~\cite{deffnerQuantumSpeedLimits2017} and more recent classical speed limits~\cite{shanahanQuantumSpeedLimits2018,okuyamaQuantumSpeedLimit2018,das2021density,*das2021speed}, the central problem of thermodynamic speed limits is to determine the minimum time to transition between two physical states or the minimum time for an observable to change by a fixed amount~\cite{JRGReview2022}. 
Despite being classical, some thermodynamic speed limits have been discovered~\cite{nicholsonTimeInformationUncertainty2020} that can be combined with the Mandelstam-Tamm time-energy uncertainty relation~\cite{mandelstamUncertaintyRelationEnergy1991} to bound dynamical observables of open quantum systems~\cite{garcia2021unifying}.
While fluctuation theorems~\cite{jarzynski1997nonequilibrium,crooks1999entropy,esposito2010three,rao2018detailed} and thermodynamic uncertainty relations~\cite{barato2015thermodynamic,gingrichDissipationBoundsAll2016,di2018kinetic,horowitz2020thermodynamic,neri2020second,dechant2021continuous,kolchinsky2021work,Godec2021Thermodynamic,skinner2021improved} bound the \textit{value} of thermodynamic observables and their fluctuations at specific points in time, thermodynamic speed limits instead bound the \textit{timescales} associated with the observable.
\jrg{Speed limits have been applied to various stochastic dynamics with continuous and discrete state spaces~\cite{aurell2012refined,shiraishiSpeedLimitClassical2018,hasegawa2019uncertainty,dechant2019thermodynamic,itoStochasticTimeEvolution2020,plata2020finite,falasco2020dissipation,nicholsonTimeInformationUncertainty2020,van2021geometrical,hamazaki2021speed,Dechant2021Improving,tasnim2021thermodynamic,falasco2021beyond,nicholsonThermodynamicSpeedLimits2021}, and unlike early bounds in finite-time thermodynamics~\cite{salamonThermodynamicLengthDissipated1983,salamonLengthStatisticalThermodynamics1985,feldmannThermodynamicLengthsIntrinsic1985,fairen1982thermodynamic}, do not rely on the system being close to equilibrium or transforming slowly.}
However, these bounds are largely on thermodynamic costs, such as the entropy production and energy wasted as heat, leaving open an important question: Are there physical bounds on the timescales associated with mechanical work? 

In this Letter, we establish upper and lower integral bounds on the minimum time to transfer energy to or from a system in the form of mechanical work.
On short timescales, \jrg{we show that} these reduce to bounds on the intrinsic speed of work set by the intrinsic speeds of heat and internal energy, establishing a ``first law of thermodynamic speeds''.
Deriving these \jrg{differential speeds from externally measured timescales} gives a rigorous definition of intrinsic speeds of both thermal and mechanical observables.
These bounds do not depend on the choice of stochastic dynamics \jrg{and relate the timescales defined in other speed limits on intrinsic timescales. W}e demonstrate them using numerical simulations of physical setups that are experimentally realizable.
Using these numerical examples, we also explain how to use empirical data and set speed limits that predict the minimum time to generate a needed amount of work.

\textit{Minimum time to generate an amount of work.---}Consider a microscopic physical system undergoing a, potentially nonstationary, stochastic process over a time $t_0\leq t\leq t_0+T$ and generating work, $\W(T)$.
What is the \textit{minimum} time, $T_{\!\scriptscriptstyle\mathcal{W}}$, it will take for the system to produce or consume a net amount of work $|\W(T_{\!\scriptscriptstyle\mathcal{W}})|=|\int_{t_0}^{t_0+T_{\!\scriptscriptstyle\mathcal{W}}}\Wdot(t) dt|$? To answer this question, we recognize that the work is bounded from above by the \textit{absolute work}, $\overline{\W(T_{\!\scriptscriptstyle\mathcal{W}})}$~\footnote{Throughout, $\overline{\mathcal{O}(t)}$ denotes the integrated absolute value of the observable $\overline{\mathcal{O}(t)}=\int_{t_0}^{t_0+t}|\dot{\mathcal{O}}(t')|dt'$.}:
\begin{align}
|\W(T_{\!\scriptscriptstyle\mathcal{W}})|
&\leq \overline{\W\left[T_{\!\scriptscriptstyle\mathcal{W}}(c)\right]} :=\int_{t_0}^{t_0+T_{{\!\scriptscriptstyle\mathcal{W}}}(c)}|\Wdot(t)| {d}  t = c.
  \label{eqIntegralT_Wc}
\end{align}
The last equality means that any process exchanging energy as work at a ``speed'', $|\dot{\W}(t)|$, for a time, $T_{\!\scriptscriptstyle\mathcal{W}}$, traverses a ``distance'', $c$.
From this perspective, the absolute value $|\dot{\W}(t)|$ ensures the time and the speed have positive values.
To put this interpretation on firmer ground, the rate of change in the time $T_{\!\scriptscriptstyle\mathcal{W}}$ with respect to the distance,
\begin{align}
  T'_{\!\scriptscriptstyle\mathcal{W}}(c):=\frac{ {d}  T_{\!\scriptscriptstyle\mathcal{W}}(c)}{ {d}   c} = \frac{1}{|\Wdot\left[T_{\!\scriptscriptstyle\mathcal{W}}(c)\right]|},
  \label{eqIntegralT_Wc2}
\end{align}
is determined by the inverse-function theorem~\cite{clarke1976inverse} when $\W$ is continuously differentiable around $T_{\!\scriptscriptstyle\mathcal{W}}$ and $\Wdot$ is non-zero. If these conditions \jrg{are not} satisfied, this relation must be determined in separate time intervals \jrg{where the conditions hold}. 
\jrg{Similar relations follow if we instead start with the heat or internal energy.}

Now, any exchange of energy as work will depend on the internal energy $\dot{U}$ and heat $\Qdot$ through the (stochastic) first law of thermodynamics, $\Wdot=\dot{U}-\Qdot$~\cite{callenThermodynamicsIntroductionThermostatistics1985}.
So, applying the triangle inequality~\cite{abramowitzHandbookMathematicalFunctions1964},
\begin{eqnarray}
  ||\dot{ U}(t)|-|\Qdot(t)|| \leq |\Wdot(t)| &\leq |\dot{U}(t)|+|\Qdot(t) |,
  \label{DotUDotQDotW}
\end{eqnarray}
leads to our first main result: 
\begin{align} 
  \frac{1}{|\dot{U}[T_{\!\scriptscriptstyle\mathcal{W}}]|+|\dot{\mathcal{Q}}[T_{\!\scriptscriptstyle\mathcal{W}}]|}\leq T^\prime_{\!\scriptscriptstyle\mathcal{W}}(c) \leq\frac{1}{||\dot{U}[T_{\!\scriptscriptstyle\mathcal{W}}]|-|\dot{\mathcal{Q}}[T_{\!\scriptscriptstyle\mathcal{W}}]||}. 
  \label{EqIntegralUncertaintyW}
\end{align}
The temporal rate of the internal energy, $\dot{U}$, and the heat flux, $\Qdot$, evaluated at the time $T_{\!\scriptscriptstyle\mathcal{W}}(c)$ set bounds on the timescale for the generation of the accumulated work.
Solving Eq.~\eqref{EqIntegralUncertaintyW}, analytically or numerically,
yields bounds on the minimum time, $T_{\!\scriptscriptstyle\mathcal{W}}$, for the work, $|\W|$, done on or by the system.
That is, the integrated bounds are on the time $T_{\!\scriptscriptstyle\mathcal{W}}$ to exchange an amount of absolute work, $c$.
The initial condition for the differential inequalities can be any pair $\overline{\W(T_{\!\scriptscriptstyle\mathcal{W}})}$ and $T_{\!\scriptscriptstyle\mathcal{W}}$. \jrg{In principle, the initial value could be chosen by the experimentalist.} 
The bounds set by Eq.~\eqref{EqIntegralUncertaintyW} are physical limits on the speed time $T_{\!\scriptscriptstyle\mathcal{W}}^\prime$.

\jrg{One can readily derive versions of Eq.~\eqref{EqIntegralUncertaintyW} 
for the time to dissipate an amount of energy as heat or cause a change in internal energy.}
\jrg{These relations could be used in situations} when the exact rates setting the bounds are known from a model or can be estimated from empirical data~\footnote{For empirical data, if $|\dot{U}(t)|$, $|\Qdot(t)|$ or $\overline{\W(t)}$ cannot be fit with smooth functions for the entire duration of measurement, the solution can be obtained for segments of time.
For example, if $\dot{\Q}$ needs two different fits when $t'\in(0,t/2)$ and $t'\in(t/2,t)$, then bounds are generated independently for $T_{\!\scriptscriptstyle\mathcal{W}}<t/2$ and $t/2<T_{\!\scriptscriptstyle\mathcal{W}}<t$, for any value of $\overline{\W}$ achievable in these time segments.
Segmenting time intervals is also necessary when $|\dot{ U}|=|\Qdot|$.}. \jrg{An advantage of these relations is that they allow for the estimation of timescales from measured data.}

\begin{figure}[t!] 
	\centering
\includegraphics[width=1.0\linewidth]{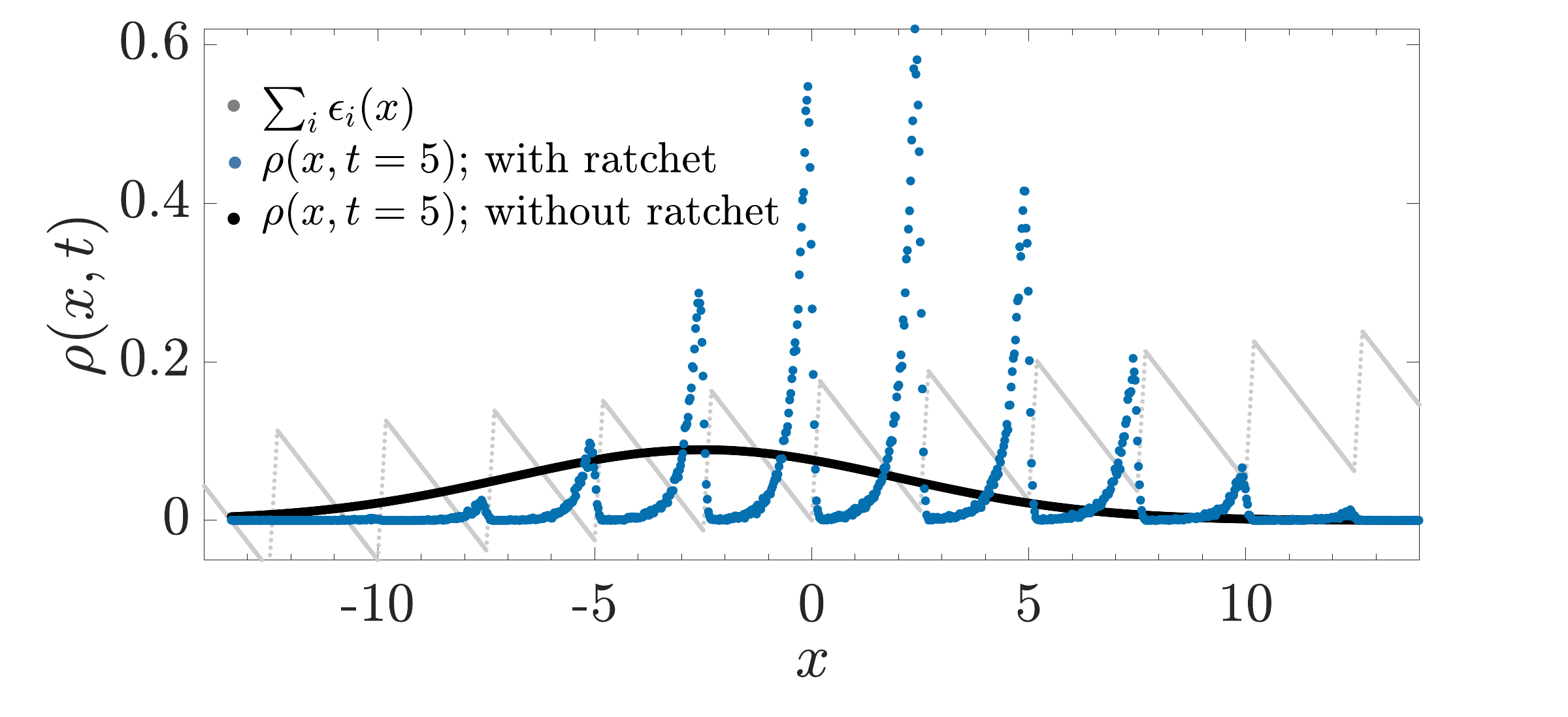}%
\vspace{-.2in}
\caption{\label{Fig1}{\footnotesize{\textit{A flashing Brownian ratchet converts thermal fluctuations into mechanical work.}---Periodically applying an asymmetric potential does work by raising particles against a constant force. 
The externally applied potential, $\sum_i\epsilon_i(x)=\epsilon_{{\text{off}}}+\epsilon(x)$, is shown in gray. It consists of the sawtooth-shaped flashing ratchet $\epsilon(x)$ (for simulation details, see App.~\ref{SecSimulationDetails}) and a time-independent component $\epsilon_{{\text{off}}}=gx$ with a constant $g$. The sawtooth potential is turned on and off periodically. The probability density function of particle positions, $\rho(x,t=5)$, obtained from simulations moves to the right as time increases as a result of thermal noise and the flashing ratchet (blue circles). The position density of particles only subject to the static potential $\epsilon_{{\text{off}}}$ drifts to the left (black).}
}}
\end{figure}

\textit{Flashing ratchet.---} \jrg{To illustrate Eq.~\eqref{EqIntegralUncertaintyW} and the ability to relate a timescale to other fluxes of energy, we consider the flashing ratchet as} a system performing work~\cite{astumian1997thermodynamics}.
\jrg{The system consists of a collection of particles along a spatial coordinate $x$ that draw thermal energy from a heat bath and leverage an asymmetric, periodically-flashing potential to do work against a constant external force, $\epsilon_{\text{off}}=gx$. For our numerical examples, $g=0.5$.} 
\jrg{Without the flashing potential, the ensemble of Brownian particles simply drifts gradually to the left (decreasing $x$) under the static force~\cite{astumian1997thermodynamics}. A representative \jrg{probability density function} is shown in Fig.~\ref{Fig1}.}

\jrg{During intermittent periods of duration $t_{\text{off}}$, the external potential is $\epsilon_{\text{off}}=gx$. 
For periods of duration $t_{\text{on}}$, there is an additional sawtooth potential: $\epsilon(x)=\text{mod}(x,L)\left[\frac{U_0}{\alpha L}R_1(x)-\frac{U_0}{L-\alpha L}R_2(x)\right]+nR_2(x)$, where $U_0,\alpha,n$ and $L$ are positive constants, and $R_1(x)$, $R_2(x)$ are spatially-periodic indicator functions~(see App.~\ref{SecSimulationDetails}, for simulation details) 
When the ratchet potential is \textit{on}, particles can fall into local wells. Because of the asymmetry of these wells, particles move to the right (increasing $x$) and against the direction of the constant force. 
Over time, the particle packet thermally expands, so this motion combined with the asymmetric shape of the wells causes a net movement of the position probability density function against the constant force. That is, work is done on the particles.}

\begin{figure}[t!]
\centering
\begin{tikzpicture}
\node (image) at (0.05,0.25) {
  \includegraphics[width=1.0\linewidth]{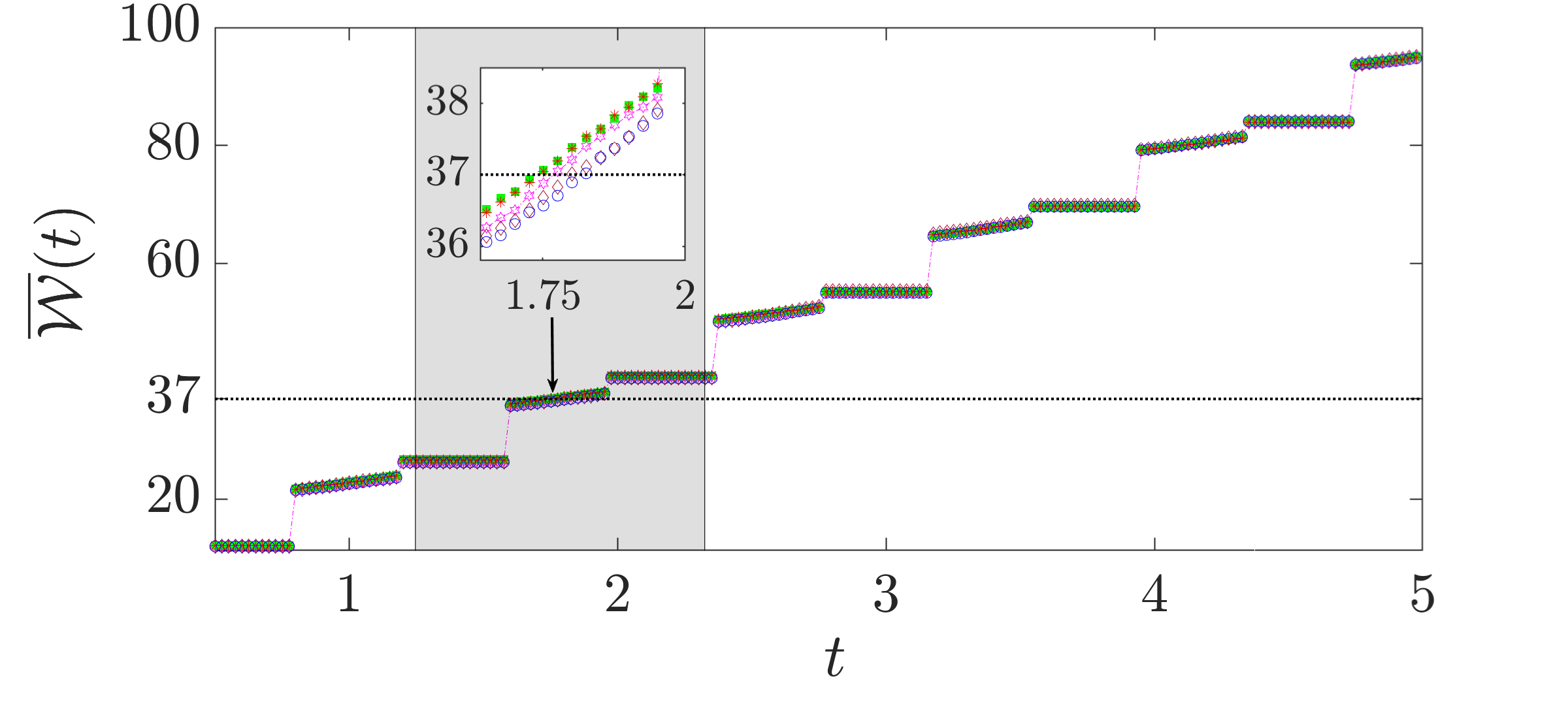}
};
\node (image) at (0,-5.75) {
 \includegraphics[width=1.0\linewidth]{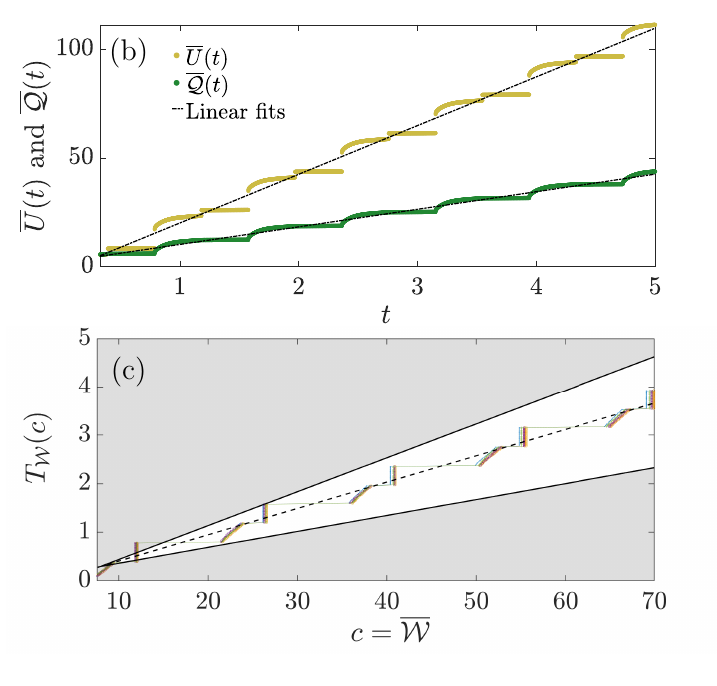} 
};
\node[text=black] (a) at (-2.75,1.7){\large{(a)}};
\end{tikzpicture}
\vspace{-.2in}
\caption{{\footnotesize{\jrg{\textit{Extracting heat and energy rates for bounds on the minimum time to produce work.}---(a) The absolute-work $\overline{\W(t)}$ in Eq.~\eqref{eqIntegralT_Wc} grows with time in five independent simulations of the Brownian ratchet (App.~\ref{SecSimulationDetails}) with 100,000 particles (colored symbols). Inset: magnification of the region between $t=1.5$ and $2$ with $t_{{\text{on}}}=t_{{\text{off}}}=0.393$. 
For an arbitrarily chosen value $\overline{\W}=37$ (dashed line), solving Eq.~\eqref{EqIntegralUncertaintyW} gives upper/lower bounds (gray) on the minimum time $T_{\!\scriptscriptstyle\mathcal{W}}(37)$. (b) 
The ensemble averaged absolute energy $\overline{U(t)}$ (yellow) and absolute heat $\overline{\Q(t)}$ (green) as function of time. Data and linear fits (dashed black lines) of one simulation. 
(c) Upper and lower bounds (solid black lines) on $T_{\!\scriptscriptstyle\mathcal{W}}(c)$ as function of $c=\overline{\W}$. Simulation data is from $|\dot{U}(t)|$ and $|\Qdot(t)|$ via Eq.~\eqref{EqIntegralUncertaintyW} in (b). 
These bounds nicely confine the results of $T_{\!\scriptscriptstyle\mathcal{W}}(c)$, from all five simulations (colored symbols). A dashed black line shows the linear fit to $T_{\!\scriptscriptstyle\mathcal{W}}(c)$ from one of these simulations.}}}} 
\label{Fig2}
\end{figure}

\jrg{The absolute work done by the flashing ratchet on the particles is a stochastic quantity.
For a collection of independent particles, the average value is an increasing function of time, $t$.
Figure~\ref{Fig2}(a) shows the absolute-work done in simulating the Langevin dynamics (App.~\ref{SecSimulationDetails}) of $10^5$ particles.
The color distinguishes five different realizations; in each, the particles start at rest from $x=0$.
The ratchet is \textit{off} for a duration $t_\text{off}$ to allow particles to diffuse and drift under the constant force.
The discontinuous jumps in the value of $|\W(t)|$ occur when the ratchet potential is switched from \textit{off} to \textit{on} or vice-versa.}

\jrg{The time $T_{\!\scriptscriptstyle\mathcal{W}}$ is the minimum time to accomplish an amount of work in absolute value; that is, it is the first-passage time of the random process $|\W(t)|$.
In Fig.~\ref{Fig2}(a), we also show the bounds on $T_{\!\scriptscriptstyle\mathcal{W}}(37)$ for $c:=\overline{\W(t)} = 37$.
They are marked by the boundaries of the gray shaded area.
Evaluating the bounds on $T_{\!\scriptscriptstyle\mathcal{W}}(c)$ for any value of $c$ requires the rates $\dot{U}(t)$ and $\Qdot(t)$.}
For the ratchet, we measure the change in internal energy from $\dot{U} = d_t\langle\sum_i\epsilon_i(x,t)\rangle$ at time $t$, where the sum is over the static, $\epsilon_{{\text{off}}}$, and flashing, $\epsilon(x)$, potentials shown in Fig.~\ref{Fig1}(a).
We measure the heat via $\Qdot=-\langle [\sum_i\epsilon_i(x,t)]\dot{I}(x,t)\rangle$~\cite{nicholsonTimeInformationUncertainty2020}, where $I(x,t):=-\ln P(x,t)$ and $P(x,t)$ is the probability  distribution  
of the particles' position at time $t$.
The work follows from the first law.

\jrg{Experimentally, the heat and internal energy can \jrg{sometimes} be modeled as linear functions.
For example, linear fits are used to extract the heat and work in the optical manipulation of colloidal particles~\cite{band1982finite,gomez2015non,martinez2017colloidal}.
For linear functions, the exact solution to Eq.~\eqref{EqIntegralUncertaintyW} is $[1/(a+b)]c+\tilde{T}_0\leq T_{\!\scriptscriptstyle\mathcal{W}}(c)\leq|1/(a-b)|c+T_0$, given $|\dot{U}|=a>0$ and $|\Qdot|=b>0$ with constants $\tilde{T}_0$ and $T_0$.
To confirm Eq.~\eqref{EqIntegralUncertaintyW}, we fit our simulation data (App.~\ref{SecSimulationDetails}) to a linear model for $\overline{U(t)}$ and $\overline{\Q(t)}$,  Fig.~\ref{Fig2}(b). From these linear fits, we find the rates of heat and work from the slopes: $\overline{U(t)}=22.41t - 2.47$ and $\overline{\Q(t)}= 8.089t + 1.988$.
Substituting these rates into Eq.~\eqref{EqIntegralUncertaintyW}, we solve the differential inequalities with the initial condition $T_{\!\scriptscriptstyle\mathcal{W}}(7.9)=0.293$ [solid black lines in Fig.~\ref{Fig2}(c)].}

\jrg{The bounds on $T_{\!\scriptscriptstyle\mathcal{W}}(c)$ that follow from this analysis match simulations well for $c\geq7.9$. 
The choice of initial condition is arbitrary; here, the choice matches a point on the best linear fit to $T_{\!\scriptscriptstyle\mathcal{W}}(c)$ from the one of the ratchet simulations.} 
\jrg{The choice of a linear model and this initial condition do cause outliers at small $c$ values. These fits are meant as an illustration, as the linear fits in Fig.~\ref{Fig2}(b) do not represent the instantaneous jumps in $U$ and $\Q$ in the ``on/off'' switching of the sawtooth potential. 
Other choices for the intercept are possible, such as $(c,T_{\!\scriptscriptstyle\mathcal{W}})=(0,0)$. 
Generally, Eq.~\eqref{EqIntegralUncertaintyW} leaves us freedom to choose the initial condition for the fitting, however it ensures that from the $\dot{U}$ and $\Qdot$ data we can obtain bounds on $T_{\!\scriptscriptstyle\mathcal{W}}$ whose tightness increases with the quality of the numerical fit.  
If the fitting function is precise, or if the energy and heat rates are known exactly (as in a model), one of the bounds will saturate.}

\medskip 
\textit{Intrinsic speeds and differential speed limits.---}The bounds in Eq.~\eqref{EqIntegralUncertaintyW} \jrg{for the extrinsic time raise the question of their connection to other speed limits on the intrinsic timescale of} observables.
Previous ``differential speed limits'' were given in terms of \textit{intrinsic speeds}, speeds defined through heuristic arguments and analogy with quantum mechanical speeds~\cite{nicholsonTimeInformationUncertainty2020,hasegawa2019uncertainty,itoStochasticTimeEvolution2020,hasegawa2021thermodynamic,JRGReview2022}.

\jrg{To reconcile the use of intrinsic and extrinsic times in thermodynamic speed limits, we establish Eq.~\eqref{eqIntegralT_Wc2} as the formal definition of an intrinsic speed.
To see this connection, we can consider smoothly varying path functions, $\mathcal{X}$, and state functions, $\langle X\rangle$.
Take a time interval where their rates, $\dot{\mathcal{X}}$ and $d_t\langle X\rangle$, do not change sign. Sufficiently short time intervals will naturally satisfy this condition, and the magnitude of the absolute integrated values will also not change significantly. Then, Eq.~\eqref{eqIntegralT_Wc} becomes:
\begin{equation}
\begin{aligned}
&\left|\int_{t_{0}}^{t_{0}+\speedtimex}\dot{X}(t)  dt\right| =  \int_{t_{0}}^{t_{0}+\speedtimex}|\dot{X}(t)| dt =: {\delta |X|}\approx\speedtimex|\dot{X}(t_0)|\\ & \left|\int_{t_{0}}^{t_{0}+\speedtimemcx}\dot{\mathcal{X}}(t)  dt\right| =  \int_{t_{0}}^{t_{0}+\speedtimemcx}|\dot{\mathcal{X}}(t)| dt =: {\delta |\mathcal{X}|}\approx\speedtimemcx|\dot{\mathcal{X}}(t_0)|.
\label{EqDefinitions1}
\end{aligned}
\end{equation}
\jrg{The equalities on the left-hand side stem from the assumption that the observable rates do not change sign in the time interval. When these rates are zero over the entire interval, the intrinsic time diverges and the observable would take an infinite amount of time to grow by any finite amount. Physically, this situation occurs at steady-states, where one would naturally expect a vanishing speed.}
The approximations on the right-hand side of Eq.~\eqref{EqDefinitions1} hold when $\dot{X}$ or $\dot{\mathcal{X}}$ are constant, when the process can be linearized, or when $\delta |X|$ and $\delta |\mathcal{X}|$ are sufficiently small.
They also relate a distance to a velocity and intrinsic timescale, and this analogy motivates the more general definition~\cite{JRGReview2022} of the intrinsic speeds.}

\jrg{Equation~\eqref{EqDefinitions1} leaves open the choice of distance $\delta|X|$ or $\delta|\mathcal{X}|$ in the definition of the intrinsic timescale.
Previous definitions~\cite{nicholsonTimeInformationUncertainty2020,itoStochasticTimeEvolution2020} chose the time that it takes for an observable to change by one standard deviation, $\Delta X$, at $t_0$. This definition is analogous to the timescale used in the Mandelstam-Tamm version of the time-energy uncertainty relation~\cite{Messiah61}. In stochastic thermodynamics, this definition leads to bounds on intrinsic speeds using the product of $\Delta X$ and variance of the Fisher information~\cite{nicholsonTimeInformationUncertainty2020,itoStochasticTimeEvolution2020}. 
With this choice:
$\speedtimex^{-1} = \left|d_t\langle X\rangle\right|/\Delta X$ and $\speedtimemcx^{-1} =  |\mathcal{\dot{X}}|/\Delta X$.
These definitions of speed are time-dependent and depend on the evolution of $\langle X\rangle$ and $\mathcal{X}$, even when they are nonlinear functions of time. We suppress this time dependence in our notation.}
\jrg{We can connect these intrinsic timescales to the extrinsic time, $T_X$, defined from $\overline{X(T_X)}=\int_{t_0}^{t_0+T_X}|\dot{X}(t)|dt$ in Eq.~\eqref{eqIntegralT_Wc}. The extrinsic timescales converge to their intrinsic timescale counterparts when $\overline{X}\rightarrow\delta|X|\rightarrow0$.} 
In this limit, the inequality $|X(T_X)|\leq\overline{X(T_X)}$ becomes an equality.
A similar logic holds for the extrinsic time of path functions, $T_\mathcal{X}$. 

\jrg{So far, intrinsic times have been defined for the internal energy and heat, but not mechanical work. 
Consequently, the definition of the intrinsic speed for work above completes the family of stochastic thermodynamics speeds~\cite{nicholsonTimeInformationUncertainty2020,JRGReview2022}.
Still open, though, is how the first law of stochastic thermodynamics constrains the relation between these intrinsic thermodynamics speeds.}
We would anticipate upper and lower bounds on the intrinsic speeds similar to Eq.~\eqref{EqIntegralUncertaintyW}.
Notice that we can divide the time series of all the observables into segments of equal, short time intervals in which the derivatives are approximately constant. 
Provided the associated speeds are small and there are no abrupt changes, the first law of stochastic thermodynamics and the triangle inequality in Eq.~\eqref{DotUDotQDotW} together yield $\left|| {\delta}  U|-| {\delta}  \Q|\right| \leq | {\delta}  \W| \leq | {\delta} U |+| {\delta} \Q |$. For the intrinsic timescale of work,
\begin{equation}
\left|f_{\!\scriptscriptstyle U}\speedtimeenergy - f_{\!\scriptscriptstyle \Q}\speedtimeheat\right| \leq  \speedtimework
  \leq f_{\!\scriptscriptstyle U} \speedtimeenergy + {f}_{\!\scriptscriptstyle \Q} \speedtimeheat, 
\label{DiffFirstLaw2} 
\end{equation}
the bounds depend on $f_{\!\scriptscriptstyle U} = |\dot{ U}|/|\Wdot|$ and ${f}_{\!\scriptscriptstyle \Q} =|\Qdot|/|\Wdot|\leq 1+{f}_{\!\scriptscriptstyle U}$, which determine the relative contributions of the internal energy and heat timescales. 
The intrinsic timescales of first law quantities are then related with a combination of any two speeds bounding the third.

\textit{\jrg{First law of intrinsic thermodynamic speeds.}---}\jrg{We can derive more direct relationships between} the intrinsic speeds. Consider the time it takes for the energy, work, and heat to change by some  fixed, equal small ``distance'' $\tilde{c}\geq 0$ during a physical transformation of the system: ${\delta |\W|}\approx \Wdot\speedtimework=\tilde{c}$, ${\delta |\Q|}\approx\Qdot \speedtimeheat=\tilde{c}$, and ${\delta |U|}\approx \dot{U}\speedtimeenergy=\tilde{c}$. 
\begin{figure}[t]
\centering
\includegraphics[width=1.0\linewidth]{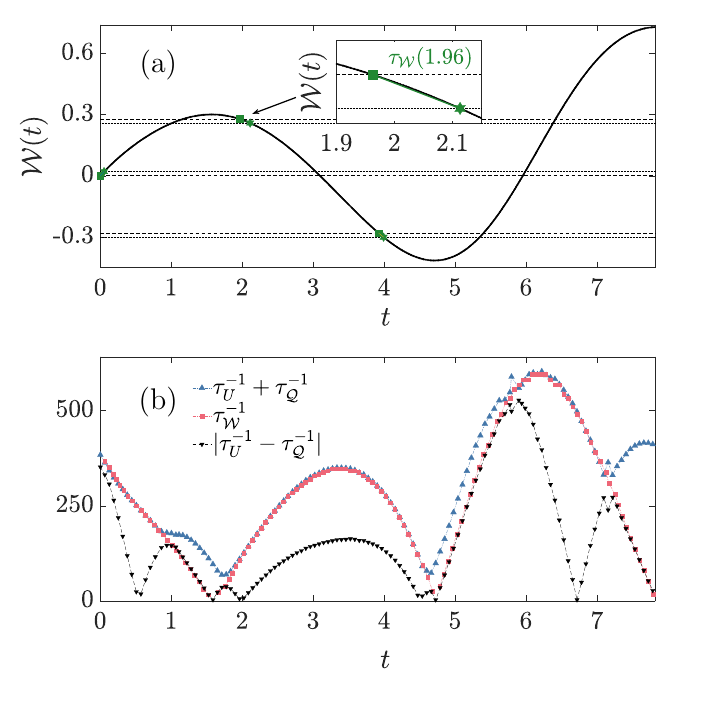} 
\vspace{-.2in}
\caption{\footnotesize{\textit{Demonstration of first law of intrinsic stochastic thermodynamic speeds.}---(a) The work $\W(t)$ (black line) done on a system of Brownian particles (for simulation details, see App.~\ref{SubSecDiffusion})  as function of time $t$. Three intervals of $|\delta\W|=0.02$ (between dash and dash-dot black lines) mark three of the locations where we measure $\speedtimework$ (green squares/stars mark the start/end of each interval). 
Linear fits in these intervals give $|\delta\W|\approx \speedtimework |\Wdot|$. The inset shows a magnification of one of these $\tau$ intervals and the linear fit to $\W$ in that interval. (b) The relation between $\speedtimework^{-1}$ (red squares) and the combinations of $\speedtimeenergy^{-1}$, $\speedtimeheat^{-1}$ for a fixed distance $|\delta\W|=0.01$. These simulation results confirm the bounds on the intrinsic speed of work in Eq.~\eqref{eq:firstlawspeeds2}. At times where $|\dot{\W}|\neq 0$, the upper bound (blue triangles) is saturated, and at other times the lower bound (black triangles), depending on the signs of $\Wdot$, $\dot{U}$, and $\Qdot$.}}
\label{Fig3} 
\end{figure} 
With fixed distances, we can recast Eq.~\eqref{DotUDotQDotW} entirely in terms of the intrinsic speeds, $\speedtimework^{-1}$, $\speedtimeheat^{-1}$, $\speedtimeenergy^{-1}$.
Using $\Wdot\approx{\delta |\W|}/\speedtimework$, $
  {\delta |U|}/{\delta |\W|} =1,
  \Hquad
  \mbox{and} 
  \Hquad
  {\delta |\Q|}/{\delta |\W|} = 1
  $, 
we find:
\begin{equation}
  |\speedtimeenergy^{-1} - \speedtimeheat^{-1}| \leq \speedtimework^{-1} \leq \speedtimeenergy^{-1} + \speedtimeheat^{-1}.
\label{eq:firstlawspeeds2}
\end{equation}
Similarly, the heat or internal energy are bounded: $|\speedtimeenergy^{-1} - \speedtimework^{-1}| \leq \speedtimeheat^{-1} \leq \speedtimeenergy^{-1} + \speedtimework^{-1}$ and $|\speedtimework^{-1} - \speedtimeheat^{-1}| \leq \speedtimeenergy^{-1} \leq \speedtimework^{-1} + \speedtimeheat^{-1}$. These expressions become equalities when one of the rates, $\Wdot$, $\Qdot$, or $\dot{U}\rightarrow 0$. For example, the intrinsic speed for work $\speedtimework^{-1}\to 0$ when $\Wdot\rightarrow 0$.
\jrg{Since they derive from the stochastic first law of thermodynamics, these results represent what we call the first law of intrinsic thermodynamic speeds. As a corollary to the stochastic first law: \textit{The intrinsic speed of a composite observable is upper (lower) bounded by the sum (difference) of the speeds of its components.}}

\textit{Driven particles in a harmonic trap.---}To demonstrate the intrinsic speed limits, we consider a collection of particles hovering in a heat bath above a flat surface, which exerts a force whose amplitude depends on the particles' height above the surface and time.
Here, work is involved in changing the potential shape with or against the direction of the packet's thermal expansion.
Experimentally, this system can be realized by optically trapped colloidal particles~\cite{astumian1997thermodynamics,grier2006holographic,martinez2017colloidal}.
We model it with a unidimensional, overdamped Langevin equation (App.~\ref{SubSecDiffusion}).
The particles start from rest at $x(0)=0$. 
Their spatial coordinate, $x(t)$, evolves via $\dot{x}(t)=\mu^{-1} F(x,t)+\sqrt{2D}\Gamma(t)$ with damping $\mu>0$, diffusion constant $D=k_BT/\mu$, and temperature $T>0$.
We set $\mu=1$ and $k_BT=1.25$.
The deterministic force $F(x,t)=[-0.2x+0.5\sin(x)]\times[1+0.75\sin(t)]$ originates from a time-dependent potential $\epsilon(x,t)$. 
The $\delta$-correlated white Gaussian noise, $\Gamma(t)$, has zero mean.
Since the modulation of the external force is continuous in time, we measure the work 
via $\Wdot=\langle\dot{\epsilon}(x,t)\rangle$~\cite{nicholsonTimeInformationUncertainty2020} and heat/energy as in the ratchet. 

Solving the master equation associated with these stochastic dynamics~\cite{barkai2014area,holubec2019physically} (see App.~\ref{SubSecDiffusion})
, we obtain the position distribution and observables for $t\in [0\rightarrow 5\pi/2]$.
The sinusoidal modulation of the potential causes the particle packet to flatten and sharpen periodically. 
Work varies over time as the external force varies. Figure~\ref{Fig3}(a) shows three representative intervals of $|\delta\W|=0.02$
setting the value of $\speedtimework(t)$. 

Figure~\ref{Fig3}(b) confirms the \jrg{relation between intrinsic speeds} in Eq.~\eqref{eq:firstlawspeeds2}.
As for the extrinsic time, $T_{\!\scriptscriptstyle \W}$, the energy rate and heat flux put speed limits on the intrinsic speed of mechanical work, $\speedtimework$.
These limits hold for any time interval that is short enough to resolve the shape of $\W(t)$.
In the figure, we use $\delta\W=0.01$~\footnote{When measuring intrinsic times numerically from data, if the starting point for measuring $\speedtimework$ is too close to a maximum or minimum of $\W(t)$, such that $\delta\W$ is not obtained, then this time interval is undefined.
This can occur in Fig.~\ref{Fig3}(a) near $t\approx1.8$ and $4.7$.}.
Unlike the extrinsic time, Fig.~\ref{Fig2}, either the upper or lower bound is always tight (provided $\Wdot\neq 0$), even while the system is driven. 
While the total accumulated work in Eq.~\eqref{eqIntegralT_Wc} is over measurable times set by an external clock, the intrinsic speeds are determined in short time intervals.
The value of the work is determined from two components which maintain constant sign during the interval $\speedtimework$.

\label{SecDiscussion}
\textit{Discussion.---}The integral and differential bounds above are a relationship between the timescales associated with work and the speeds of heat and internal energy.
Any two observables that can be measured directly will set bounds on the third observable.
For example, in a mechanical system subject to a time-dependent potential, the internal energy is measured from the rate of change of the potential energy.
The heat flux can also be related to the rate of change in the entropy flow, $\dot{S}_e$~\cite{vandenbroeckEnsembleTrajectoryThermodynamics2015}, through $\Qdot=T\dot{S}_e$.
We can define the intrinsic time of entropy flow, $\speedtimeTSe$, from the relation: $|T\delta S_e|=|T\int_{t_0}^{t_0+\speedtimeTSe} \dot{S}_e(t)dt|$.
Using $\speedtimeTSe$ with a fixed ``distance'' $T\delta S_e=c$, Eq.~\eqref{eq:firstlawspeeds2} can now be written as $|\speedtimeenergy^{-1} - \speedtimeTSe^{-1}| \leq \speedtimework^{-1} \leq \speedtimeenergy^{-1} + \speedtimeTSe^{-1}$.
While in this letter we focus on mechanical work, the speed limit derived here applies to any linear combination 
of observables, such as the decomposition~\cite{vandenbroeckEnsembleTrajectoryThermodynamics2015} of the total entropy rate $\dot{S}=\dot{S}_i+\dot{S}_e$ into the rates of entropy flow, $\dot{S}_e$, and the entropy production in the system, $\dot{S}_i$.


Extrinsic times, like $T_{\!\scriptscriptstyle\mathcal{W}}$, are potentially useful to predict the minimal time it will take a physical, chemical, or biological system to produce a certain desired amount of work.
The bounds we have derived on these times, Eq.~\eqref{EqIntegralUncertaintyW}, led us here to the first intrinsic speed for mechanical work.
They also led us to the first formal definition of intrinsic timescales, which measure the distinguishability of observables and local fluxes of energy~\cite{nicholsonTimeInformationUncertainty2020,JRGReview2022}.
There, the Fisher information, $I_F$, sets the time it takes a probability distribution to evolve to a new statistically distinguishable state~\cite{JRGReview2022} and a speed limit $\tau^{-1}=\sqrt{I_F}$ on the intrinsic speeds of any observable with the same mathematical form as the heat: $\speedtimeheat^{-1}\leq\tau^{-1}$.
\jrg{We can use this speed limit together with the relationship between intrinsic speeds from the first law to constrain the intrinsic speed of mechanical work from above.
From Eq.~\eqref{eq:firstlawspeeds2}, we can now write $ \speedtimework^{-1} \leq \speedtimeenergy^{-1} + \tau^{-1}$.}

The formal definition of intrinsic speeds makes this result more rigorous and clearly connects intrinsic and extrinsic times.
To see how, we use the inverse function theorem~\cite{clarke1976inverse} for an observable, like heat, that is represented as a covariance, $\mathcal{A}(t)=\cov(A,\dot{{I}})$.
From $|\mathcal{A}[T_\mathcal{A}(c)]|\leq\int_{t_0}^{t_0+T_{\mathcal{A}}(c)}|\dot{\mathcal{A}}(t)|dt=c$, we can use the time-information uncertainty relation, $|\dot{\mathcal{A}}|\leq\sqrt{I_F}\Delta A$, where $\Delta A$ is the standard deviation of $A$~\cite{nicholsonTimeInformationUncertainty2020}.
Substituting gives $d_cT_\mathcal{A}(c)\geq 1/(\sqrt{I_F}\Delta A)|_{T_\mathcal{A}(c)}$.
Thus, given the Fisher information, we can obtain the first time $T_\mathcal{A}(c)$ to reach $|\mathcal{A}|=c$ for any $c>0$.
To recover the thermodynamic speed limit in Ref.~\cite{nicholsonTimeInformationUncertainty2020}, we choose $dc \approx \Delta A$, yielding $dT_{\mathcal{A}} \geq 1/\sqrt{I_F}$.
Taking $dT_\mathcal{A}=\Delta A/\dot{\mathcal{A}} = \speedtime$, we get $\speedtime \geq 1/\sqrt{I_F}$, which agrees with Ref.~\cite{nicholsonTimeInformationUncertainty2020}. 
Unlike~\cite{nicholsonTimeInformationUncertainty2020}, where $\sqrt{I_F}$ alone bounds intrinsic observable timescales, here the statistical length is evaluated at $T_{\mathcal{A}}$ and, so, depends on the observable. 


Generating work is a critical aspect of physical systems and devices, but also the dynamical functions of living systems.
Systems that are small, and driven stochastically, present particular challenges to the prediction of work that can be accomplished in a given amount of time, but also the practical design of optimal protocols in fluctuation-induced motion~\cite{brown2017toward}. 
The bounds here on the minimum time to accomplish mechanical work\jrg{, heat, or internal energy} are theoretically grounded and can be implemented in practice.
They are a relationship between stochastic observables that can be manipulated to bound an unknown quantity from others that are directly measurable.
\jrg{The extrinsic and intrinsic and speed relations here hold for systems that are both open and driven. 
By clarifying the distinction between extrinsic and intrinsic timescales, we can turn the stochastic first law of thermodynamics into a corollary relating thermodynamic speeds.}

\begin{acknowledgments}

This material is based upon work supported by the National Science Foundation
under Grant No.~1856250.
J.R.G.\ acknowledges helpful conversations with Sky Nicholson and Luis~Pedro Garc\'{i}a-Pintos.

\end{acknowledgments}

\bibliography{./references}

\appendix 

\section{Simulation details for the Brownian ratchet}
\label{SecSimulationDetails} 
The motion of a thermal particle inside a flashing Brownian ratchet is simulated using the Euler-Mayurama integration scheme (with time-step of duration $dt=0.001$) applied to the overdamped Langevin equation; $\dot{x}(t)=-\gamma^{-1}\epsilon'(x,t)+\sqrt{2D}\Gamma(t)$. Here, $\Gamma(t)$ is white Gaussian noise, with zero mean and $\delta-$function autocorrelation. As explained in the main text, the external potential $\epsilon(x,t)$ is comprised of a static part $=gx$, where $g=0.5$ (we use arbitrary units for  mass, space and time and so on, and mass is always unity without loss of generality, for real units see~\cite{astumian1997thermodynamics}), and a flashing part; $\epsilon(x)=\text{mod}(x,L)\left[\frac{U_0}{\alpha L}R_1(x)-\frac{U_0}{L-\alpha L}R_2(x)\right]+nR_2(x)$. In the latter, the function $R_1(x):=1$ if $\text{mod}(x,L)\leq(\alpha L)$ and zero otherwise, and $R_2(x):=1-R_1(x)$. These functions form the saw-tooth shape of the potential.  Using the notations in~\cite{astumian1997thermodynamics}, the other parameters defining the shape of the ratchet $U_0,L$ and $\alpha$ are $=17.5,2.5$ and $0.075$, and $n=U_0/(1-\alpha)$. The diffusion coefficient is $D=k_BT/\gamma$, where the damping $\gamma=1$ and $k_BT=2$ is the Boltzmann constant (we use $k_B\equiv1$) times the temperature $T$ of the heat bath. 

We initialize the positions of $100,000$ particles at $x=0$, and let them diffuse under the effect of the static force alone for a time duration $t_{\text{off}}=0.3929$. Then, we periodically switch the flashing part on,  for a duration $t_{\text{on}}=t_{\text{off}}$, and off for $t_{\text{off}}$, and on again until the sum of the intervals reaches $t=5$. These durations were chosen to match $t_{\text{on}}=1.1\gamma L^2/U_0$~\cite{astumian1997thermodynamics}. Importantly, during this time particles at the top of the saw-teeth have enough time to slide and get stuck inside the wells and, on the other hand, during the off times the particle packet has sufficient time to expand in both directions such that particles can reach the "territories" of more distant wells (by the time that the flashing potential turns on again). 

\section{Simulation details for particles in a time-dependent potential well}
\label{SubSecDiffusion} 
To demonstrate Eq. (8) and Fig. (3) in the main text, we used simulation results of the Langevin process  
\begin{equation} 
\dot{x}(\tau)=\frac{1}{\mu} F(x,t)+\sqrt{2D}\Gamma(\tau),  
\label{EqLangevin}
\end{equation} 
where, $\mu=1$ and $D=1.25$, $\Gamma(t)$ is similar to the random force in the Brownian ratchet example, and the deterministic force is  $F(x,t)=[-0.2x+0.5\sin(x)]\times[1+0.75\sin(t)]$. We start the dynamics with all particles set at rest on $x(0)=0$, and solve the discrete time and space master equation   
    \begin{align}
    \dot{P}(y,t) =&W_{y,y+\Delta}(t)P(y+\Delta,t)+W_{y,y-\Delta}(t)P(y-\Delta,t)\nonumber\\ 
    &-\left[W_{y+\Delta,y}(t)+W_{y-\Delta,y}(t)\right]P(y,t).    
    \label{MasterEq1} 
\end{align} 
This equation describes a discrete-space and time jump process,  analogues to the Langevin process above, with nearest-neighbour jumps to the left (-) or right (+) along a lattice $y$ embedded within $x$, with spacing $\Delta$.  The lattice can be thought of as $y=\left\lfloor x\right\rfloor \Delta$, and $P(y,t)$ is the probability of being at $y$ at time $t$, namely 
in the interval $[x,x+\Delta)$ at time $t$, after starting from $x=0$, so $\sum_y P(y,t)/\Delta\equiv\sum \rho(x,t)dx=1$, where $\rho(x,t)$ is the probability density. The spatial spacing $\Delta =0.1$. To solve Eq. \eqref{MasterEq1}, we use an Euler-Mayurama integration scheme where $\dot{P}(y,t)=\left[P(y,t+\Intd\tilde{t})-P(y,t)\right]/\Intd\tilde{t}$, and  $\Intd\tilde{t}=\Intd t\Delta^2/(2D)$ (we choose $dt=0.01$). 
 The jump probabilities are~\cite{barkai2014area
 } 
 \begin{align}
    W_{y\pm\Delta,y}(t)=\frac{1}{2}\pm\frac{F(y,t)\Delta}{4D\mu}.
    \label{JumpProbabilities}
\end{align} 
From $P(y,t)$, we obtain the probability density $\rho(x,t)$ and all the observables.

\end{document}